\documentstyle[12pt,aasms4,psfig]{article}

\newcommand{\solm}{M$_{\odot}$}

\newcommand{\COz}{$^{12}$CO (2-1) }

\newcommand{\rf}{\par\noindent\hangindent 15pt {}}
\newcommand{\appss}[2]{Ap. \& S.S.,  #1, #2.}
\newcommand{\apjj}[2]{Ap. J., #1, #2.}
\newcommand{\apjjs}[2]{Ap. J. Supp., #1, #2.}
\newcommand{\apjjl}[2]{Ap. J. (Letters), #1, #2.}
\newcommand{\asa}[2]{Astron. Astrophys., #1, #2.}
\newcommand{\arasa}[2]{Ann.Rev.Astr.Ap., #1, #2.}
\newcommand{\asas}[2]{Astron. Astrophys. Suppl., #1, #2.}

\newcommand{\mn}[2]{M.N.R.A.S., #1, #2.}
\newcommand{\ajj}[2]{A. J., #1, #2.}

\newcommand{\vol}[1]{1}

\lefthead{Schinnerer et al.}
\righthead{Molecular Gas in Seyfert Galaxies}

\begin{document}

\title{
The Molecular Gas in the Circumnuclear Region of Seyfert 
Galaxies\altaffilmark{1}}

\author{
E. Schinnerer, A. Eckart, L.J. Tacconi }

\affil{Max-Planck-Institut f\"ur extraterrestrische Physik, 85740 Garching, Germany}

\altaffiltext{1}{Based on observations carried out with the IRAM
Plateau de Bure Interferometer. IRAM is
supported by INSU/CNRS (France), MPG (Germany) and IGN (Spain).}

\begin{abstract}
Sub-arcsecond 
IRAM Plateau de Bure
mm-interferometric observations 
of the \COz line
emission in the Seyfert~1 NGC~3227 and the Seyfert~2 NGC~1068 have revealed
complex kinematic systems in the inner 100~pc to 300~pc that are not consistent
with pure circular motion in the host galaxies. Modeling of
these kinematic systems with elliptical orbits in the plane of the host galaxy
(representing gas motion in a bar potential) is a possible solution
but does not reproduce all
features observed. A better description of the complex kinematics is
achieved by circular orbits which are tilted out of the plane of the host
galaxy. This could indicate that the thin circumnuclear gas disk is
warped. In the case of NGC~1068 the warp model suggests that at a
radius of $\sim$ 70~pc, the gas disk is oriented edge-on providing
material for the obscuration of the AGN nucleus.
The position-velocity diagrams show rising rotation curves at $r \leq$
13 pc and an indication for large  enclosed masses of
$\ge$2$\times$10$^7$\solm ~
for NGC~3227 and $\ge$10$^8$\solm ~ for NGC~1068 within the central 25~pc.
\end{abstract}


\section{INTRODUCTION}

Molecular gas in the circumnuclear regions of nearby Seyfert galaxies 
can now be studied via mm-interferometry at sub-arcsecond resolution and high
sensitivity. 
In the standard unified scheme a torus of dense molecular gas and dust 
surrounding the AGN and its accretion disk (see
Peterson 1997 for an overview) is responsible for determining whether
the source is seen as a Seyfert~1 or Seyfert~2 depending only on whether 
the viewing angle
onto the central engine is blocked by the torus or not.
\\
Two questions which can be addressed by mm-interferometric
observations arise from this picture: (1) On which scales and
how much is the molecular gas contributing to the obscuration? 
(2) What is the
transport mechanism which brings the molecular gas down to the AGN?
\\
We have chosen NGC~3227 and NGC~1068 
as representative Seyfert~1 and Seyfert~2 templates to address these issues
through a detailed analysis of the molecular gas. 
An extensive description of the results for both sources will be given in two 
forthcoming papers (Schinnerer et al. 1999 and Schinnerer, Eckart,
Tacconi 1999; see also Schinnerer 1999).
NGC~3227 (D = 17.3~Mpc, group distance; Garcia 1993) is of type 1.2
(Osterbrock 1977) and 
contains a large amount of molecular gas in its central region (Meixner et
al. 1990). It has an ionization cone mapped with HST in the [O III]
line emission (Schmitt \& Kinney 1996).
\\
NGC~1068 (D = 14~Mpc; Bland-Hawthorn et al. 1997) is the archetypical 
Seyfert~2 galaxy. Antonucci \& Miller
(1985) observed emission lines widths typical for BLR lines of
Seyfert~1's in the polarized optical emission of this galaxy.
In addition to bright molecular spiral
arms at r $\sim$ 1~kpc there is also molecular gas
observed at the nuclear region (Jackson et al. 1993, Tacconi et al. 1994, 
Helfer \& Blitz 1995, Tacconi et al. 1997, Baker \& Scoville 1998). A prominent ionization
cone is traced in the [O III] line emission (Macchetto et al. 1994).

\section{OBSERVATIONS AND RESULTS}

In both galaxies the \COz line emission was observed in the winter of 1996/1997 
using the IRAM millimeter
interferometer (PdBI) on the Plateau de Bure, France, 
in its AB configuration. The five antennas were positioned
in three (NGC~3227) and two (NGC~1068) different configurations
providing 30 and 20 baselines, respectively.
The angular resolution is $\sim$ 0.6'' for NGC~3227 and $\sim$ 
0.7'' for NGC~1068. Further details of the observations will be given
in Schinnerer, Eckart, Tacconi (1999) and Schinnerer et al. (1999).
\\
\\
As shown in Fig. 1a and 1b the nuclear \COz emission in both
galaxies has a ring-like distribution.
Fig. 2 and Fig. 3 show position velocity diagrams taken along or close
to the kinematic axes of NGC~3227 and NGC~1068, respectively.
The extreme velocities (indicated by arrows) seen for the first time
in the molecular line emission of {\bf both} galaxies 
indicate rising rotation curves towards smallest radial separations and 
large enclosed masses of $\geq$ 2$\times$10$^7$ \solm ~for NGC~3227
and $\geq$ 10$^8$ \solm ~for NGC~1068 (see caption of Fig. 2 and Fig. 3).
The general gas motion in NGC~3227 can be
described by a rotating disk for radii $>$1''.
For this region we obtained a rotation curve using 
the GIPSY routine ROTCUR.
For NGC~1068 a lower limit to the rotation velocity at each
radius $\le$12'' (well within the spiral arms at r$\sim$15'')  was obtained
by averaging the extreme measured velocities on opposite sides of the nucleus
independent of position angle.
In the inner few parsecs the rotation curve was replaced by Keplerian
rotation velocities corresponding to the estimated enclosed masses.

\section{THE MODELING}

The observed position-velocity data cubes were modeled using three dimensional 
gas orbits  
and translating the motion along the line-of-sight axis onto the
spectral axis. The orbits representing gas motion are not self-intersecting 
and do not have strong cusps, since these lead to clumping, 
dissipation of kinetic energy, and therefore result in unstable orbits
(e.g. Friedli \& Martinet 1993).
Under these assumptions two ways 
to approximate the gas motions in the circumnuclear region are: 
(a) planar elliptical orbits and 
(b) tilted circular orbits. 
In case (a) the elliptical orbits resemble the two
main $x_1$- and $x_2$-families which are present in bar potentials
(see review by 
Sellwood \& Wilkinson 1993). At the position of resonances the stars flip 
from one family to
the other, whereas the gas smoothly follows this change.
The behavior can be mimicked by ellipses with changing
eccentricities and position angles (see for example Fig. 7 in Telesco
\& Decher 1988). In case (b) the tilted circular orbits form a
precessing warp in the gas disk. Such warps are quite common in HI
disks in the outer regions of galaxies (see review by Binney 1992), and 
are also observed in the accretion
disks around AGN (e.g. NGC~4258, Miyoshi et al. 1995). 
We neglect radiative transfer processes assuming that, due to the
large nuclear velocity dispersions, the bulk of the molecular gas is
not strongly effected by self-absorption.
\\
\\
The model subdivided the disk into many single (circular or elliptical) 
orbits of molecular gas.
For the modeling the inclination, position angle and shape of the
rotation curve for each host galaxy were held fixed.
Each fitting process was started at large radii and successively extended
towards the center. For each case we tried several start set-ups that
all converged to similar (best) solutions with mean deviations from
the data of less than about
10 km/s and 0.1'' for each velocity and radius in the pv-diagrams and 
10$^o$ in the position angle of the mapped structures. 
\\
To model the warp we followed the method of
Tubbs (1980; see also Quillen et al. 1992). In this approach the warp
is produced by a smooth variation of the tilt $\omega(r)$ of each
orbit relative to the plane of the galaxy and the precession angle
$\alpha(r)$. A torque acting on an orbit with a circular velocity
$v_c(r)$ introduces a precession rate
$d \alpha/dt \sim \xi  v_c/r$. After a time $\Delta$$t$ one obtains
$\alpha(r) = \xi \Omega \Delta t  + \alpha_0$. Here $\alpha_0$ is a
constant, $\Omega$=$v_c$/$r$, and $\xi$ is given by the acting torque. 
We considered for our analysis models with constant $\xi \Delta t$
and assume the molecular gas to be uniformly distributed.
\\
For the bar approach we fitted the orbital eccentricity $\epsilon (r) =
b(r)/a(r)$ ($b$ is the minor axis, $a$ is the major axis) and the
position angle $PA_{ellipse}(r)$ as curves varying smoothly with radius under 
the constraint that the orbits do not overlap with each other. 
The orbits lie in a plane and resemble velocity and density
distributions similar to the bar models of Athanassoula (1992).
\\
\\
The fitting was done on host galaxy kinematic major and minor axes
pv-diagrams and checked on the
overall intensity map and the velocity field. Here we concentrate on
the central few arcseconds and present the
results for the two galaxies by first describing the essential
properties of the pv-diagrams and then giving an outline of the best 
bar and warp models.

\subsection{NGC~3227}

For NGC~3227 two representative pv-diagrams are shown in Fig. 2. They
are taken along the major kinematic axis (PA~158$^o$) and a position angle 
of PA~40$^o$ consistent with spatial extent of the nuclear emission.
Along the major axis the drop of the rotational
velocity can be seen down to r $\sim$ 0.5''. For smaller radii
an apparent counter rotation is observed between
0.2'' $\leq r \leq$ 0.5''. For even smaller radii a second flip back
to the original sense of rotation 
is detected. These changes in 
rotation form a {\sf S}-shape in the inner 1'' of the pv-diagram. At PA~40$^o$
(close to the kinematic minor axis) a similar behavior is seen
with the exception that the change in the sense of rotation is
already occurring at $r \sim$ 0.6''. East of the dynamical nucleus 
this pv-diagram clearly shows an enhancement of emission which is
poorly reproduced by the models, since it is not axisymmetric.
\\
\\
{\it  The bar approach:}
Fig. 2 shows that we are not able to account for the observed amount of
counter rotation along the kinematic major axis and especially along PA~40$^o$
close to the kinematic minor axis. Also, in both pv-diagrams the {\sf
S}-shape in the inner 1'' is not all reproduced. Combined with the fact
that high resolution NIR observations (Schinnerer, Eckart, Tacconi
1999, Schinnerer 1999) show no evidence for a nuclear bar, this suggests
that the observed structure and {\it kinematics} of the inner 1'' are not well 
represented by motion of gas in a {\it nuclear} bar potential.
However, the circumnuclear ring ($r \approx$ 1.5'') may well be due to
an ILR of an outer structure.
\\
{\it The warp approach:}
Fig. 2 shows the fits of the best warp model to both pv-diagrams. The
model reproduces the observed {\sf S}-shaped changes in the rotational
sense in both pv-diagrams, as well as 
the observed intensity distribution in the inner
1'' (Fig. 1a). This success in reproducing {\it both} the kinematics and source
structure suggests that a warped disk, rather then an inner bar, is a
more realistic description of the molecular gas in the central 70 pc
of this galaxy.

\subsection{NGC~1068}

The kinematic major axis of NGC~1068 lies along a position angle
of PA $\sim$ 110$^o$; and the position-velocity cut along this PA
is remarkably symmetric (Fig. 3). The large scale disk emission
is observed at radii $\geq$ 2''. A large velocity
dispersion, which even rises above the rotational velocity, is seen
at $r =$ 1''. Finally we also observe emission at the systemic velocity 
which is detected at radii $\leq$ 0.9'' and in a ridge which 
connects the two high dispersion points at $r =$ 1''.
\\
\\
{\it The bar approach:}
The best bar model reproduces the overall shape of the pv-diagram.
However,
Fig. 3 shows that although the high velocity dispersion at 1'' can
be fitted, this model fails to produce the rise in velocity relative 
to the neighboring disk emission and the emission ridge. The
bar model requires highly elliptical orbits inside the ring thus
making the interpretation as an ILR of the 2~kpc stellar NIR bar
(Scoville et al. 1988) very unlikely, since it is expected
that the orbits become more circular inside the ILR (e.g. Athanassoula
1992).
We regard the bar model as a satisfying but not good fit to the data -
especially given the fact that
the nuclear stellar cluster as mapped by Thatte et
al. (1997) shows no indication for separate bar-like structure that might
induce the highly elliptical orbits. 
\\
{\it The warp approach:}
The best warp model is displayed in Fig. 3. All observed kinematic
features are reproduced by this model: 
the disk emission, the high dispersion plus rise in
velocity, the emission at the systemic velocity as well as the ridge
between the high dispersion peaks. We thus favor a warped disk over a
nuclear bar distribution to describe the gas
distribution and motion in NGC~1068 at r $\leq$ 1.5''.
\\
In this model the sudden increase in observed projected velocity as well as 
the high velocity dispersion at 1'' is due to molecular gas forming an
edge-on disk at this radial distance (see Fig. 4). Such a disk is in agreement 
with the observed orientation of the NIR/MIR polarization vectors 
(Young et al. 1996, Lumsden et al. 1999)
and the extinction band across the nuclear region 
(Catchpole \& Boksenberg 1997). The three-dimensional geometry
of the warp also naturally provides a cavity for the ionization cone in
consistency with the observed orientation (see Fig. 4).

\section{CONSEQUENCES AND IMPLICATIONS}

The warp model provides the better fit to the kinematics of the molecular gas 
in the circumnuclear regions of both studied Seyfert galaxies (as
outlined above). The bar
model might work if one allows for high streaming motions providing
further velocity modulation in the pv-diagrams. This
would, however, favor stronger nuclear bars which are {\bf not} indicated by high
resolution NIR observations (Thatte et al. 1997, Schinnerer, Eckart,
Tacconi 1999, Schinnerer 1999). 
\\
The fits to the data imply that 
even at radii as small as $r \sim$ 75 pc the gas stays in a thin disk 
with low velocity dispersion ($<$30 km/s) while the stars show an
almost spherical distribution at these distances. 
The magnitude of the torque estimated from the warp model (Sparke
1996) implies that the most likely
cause for such a warp of this thin gas layer is a torque induced by the
gas pressure of the ionization cones in both
galaxies. An important future test will be a comparision of molecular gas
kinematics in galaxies with and without ionization cones.
Alternatively, as a transient phenomenon, complexes of 
molecular clouds which do not
participate in the overall gas motion can also induce a torque on the
gas disk. 
\\
Since the observed features are symmetric with respect to the dynamical
center, a scenario in which the source structure is dominated by randomly
distributed molecular cloud complexes also appears to be very unlikely. 
Our results suggest that future theoretical studies of molecular gas
dynamics in the circumnuclear regions {\it will require} 
3 dimensional modeling. 
\\
The molecular gas being distributed in a thin, warped disk can have
important consequences for AGN obscuration (and the structure and
evolution of the NLR). Such a disk when viewed edge-on - as in the
case of NGC~1068 - can effectively hide the Seyfert nucleus from
direct view. This conclusion is supported by other recent studies of samples
of Seyfert galaxies which find that the molecular gas and dust at distances of
about 100~pc can play a significant role in the classification of Seyfert 
galaxies (Malkan et al. 1998, Cameron et al. 1993). 
\\
Our results also indicate that 
that nuclear bars are not necessarily the primary fueling mechanism
for AGN activity, since we can describe the gas distribution very well
by a uniform warped disk. This is in agreement with data presented by 
Regan \& Mulchaey 
(1999) who did not find evidence for strong bars in their Seyfert
sample. As the ionization cone is likely to cause the warp
this would then also provide the connection between the AGN 
and the inner part of its host galaxy. 
Supported by these statistical findings our results imply that the AGN
and its host galaxy are not discrete systems but are naturally linked
to each other.

\vspace*{1cm}

\acknowledgements
We like to thank the IRAM  Plateau de Bure staff for carrying out
the observations and the staff at IRAM Grenoble, especially R. Neri and
D. Downes, for their help during the data reduction. For fruitful and
stimulating discussions
we thank A. Baker, D. Downes, P. Englmaier, R. Genzel, O. Gerhard, A. Quillen,
N. Scoville and L. Sparke.

\newpage

\newpage

\begin{figure}
\psfig{file=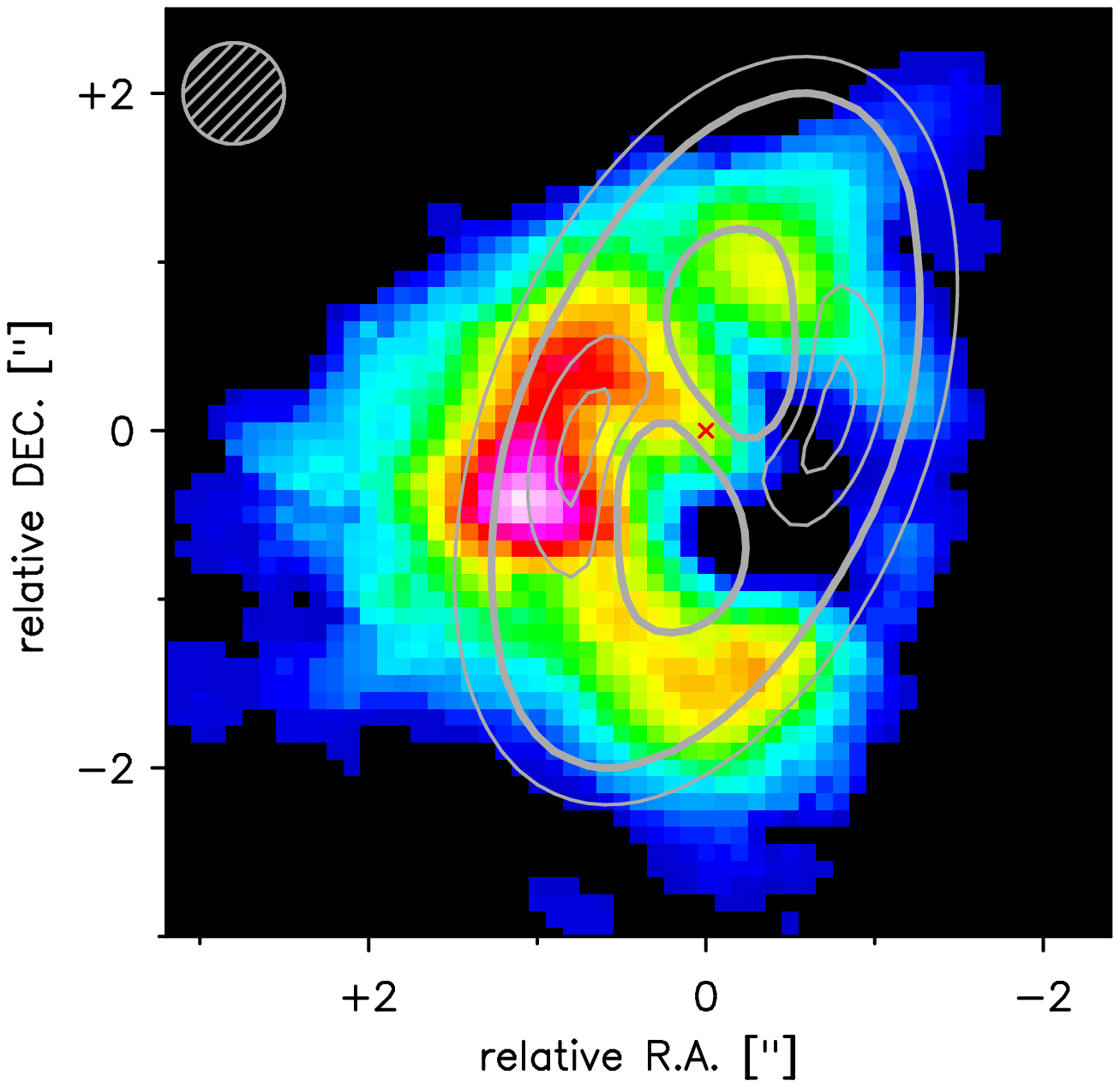,height=15.0cm,width=15.0cm,angle=0.0}
\end{figure}

\begin{figure}
\psfig{file=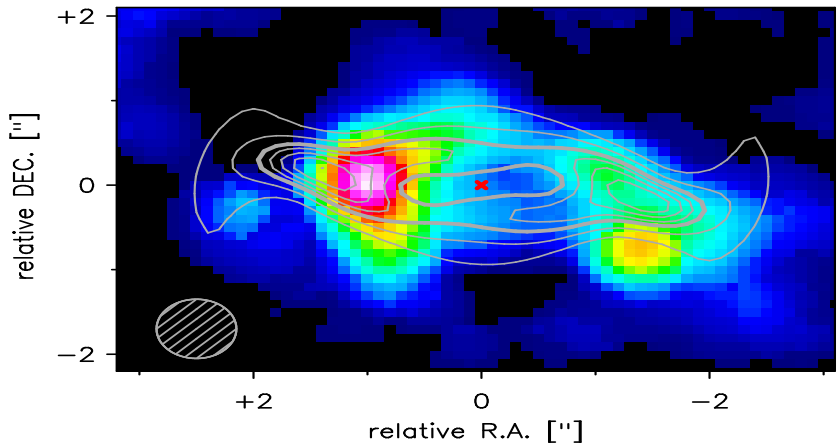,height=15.0cm,width=15.0cm,angle=0.0}

\figcaption[]{Intensity maps of the \COz line emission in NGC~3227 (a)
and NGC~1068 (b) in colour. In both galaxies the \COz emission is distributed in
ring-like features with radii $\sim$ 150~pc. In both sources nuclear line
emission is clearly detected ($>$ 3$\sigma$). The contours show the intenisty 
maps of the best warp models. Contour levels for NGC~3227 (a) are 40 \%, 65 \%
(thick), 80 \%, 90 \% and for NGC~1068 (b) 30 \%, 45 \%, 60 \%
(thick), 70 \%, 80 \%, 90 \% of the peak intensity. The thick contour lines 
outline the
circumference of the overall structures and their local minima. The
red crosses mark the positions of the fitted dynamical centers.
}
\end{figure}

\begin{figure}
\psfig{file=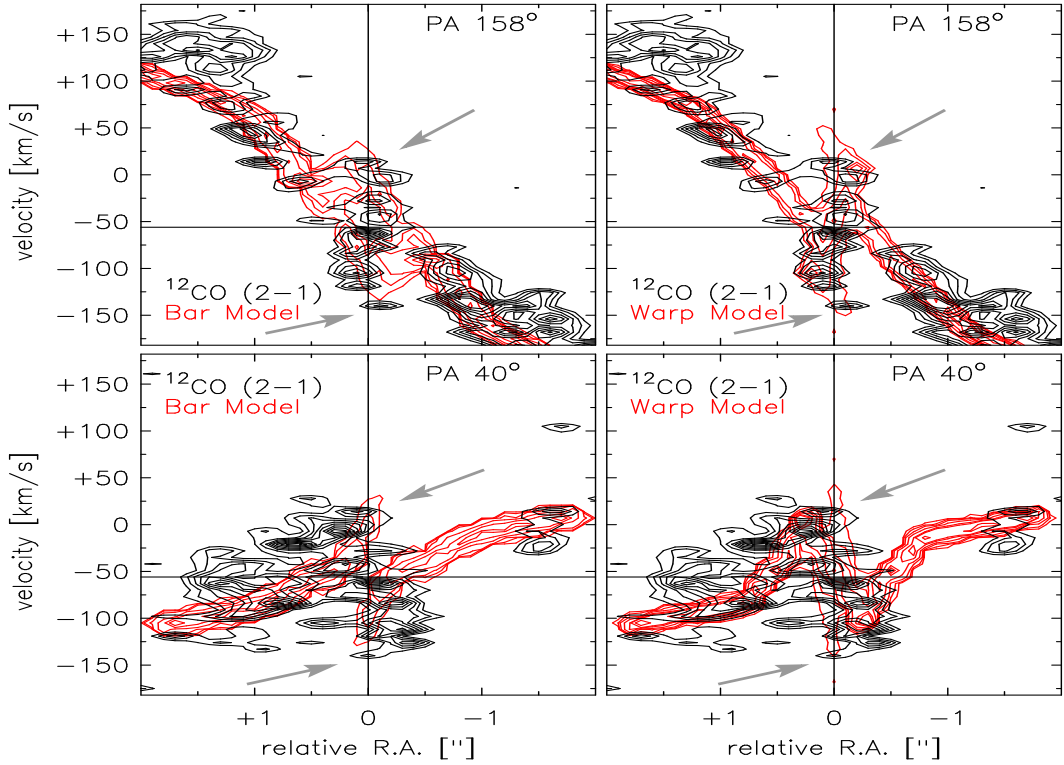,height=15cm,width=15.0cm,angle=0.0}

\figcaption[]{pv-diagrams in NGC~3227. Top: 
The data (black contours) along the kinematic major axis (PA~158$^o$) 
are shown together with the results (red contours) of the 
bar model (left) and the warp model (right). Similarly 
the data and the results along PA~40$^o$ are shown at the bottom.
In the pv-diagram at a resolution of 0.6''
the lowest contour corresponds to 3$\sigma$, with 1$\sigma$= 6.2
mJy/beam at a spectral resolution of 7 km/s.
To highlight the complex velocity behavior
the data is shown at 0.3'' (higher than the effective resolution)
close to the resolution of the model ($\sim$ 0.2'').
For the first time seen in the molecular gas emission the data shows
a rising rotation curve towards smaller radii with the extreme velocities 
(indicated by arrows)
at radial separations of $\approx$0.15'' (13~pc)
indicating an enclosed mass of $\ge$2$\times$10$^7$\solm, not correcting
for inclination effects.
This value is consistent with $\sim$10$^8$\solm ~estimated from
H$\beta$-reverberations mapping by Salamanca et al. (1994).
}
\end{figure}

\begin{figure}
\psfig{file=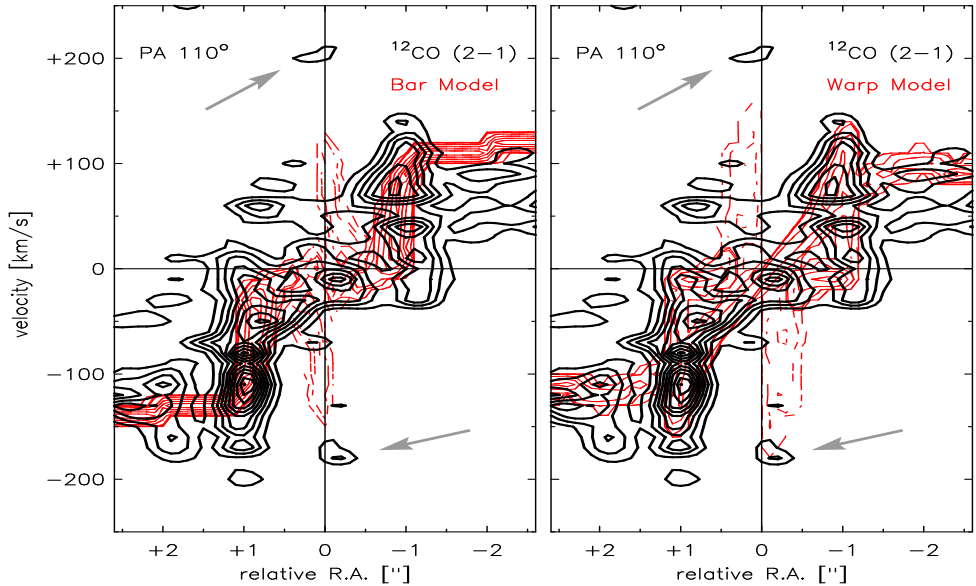,height=15cm,width=15.0cm,angle=0.0}

\figcaption[]{pv-diagrams in NGC~1068: The 
data (black contours) along PA 110$^o$ are shown together 
with the results 
(red solid contours) of the bar model (left) and the warp model (right).
In the pv-diagram at a resolution of 0.7''
the lowest contour corresponds to 3$\sigma$, with 1$\sigma$= 11
mJy/beam at a spectral resolution of 10 km/s.
To highlight the complex velocity behavior
the data is shown at 0.4'' (higher than the effective resolution)
close to the resolution of the model ($\sim$ 0.2'').
As in NGC~3227 the velocity
data at radial separations of $\approx$0.18'' (13~pc) (indicated by
arrows) indicates an enclosed mass of $^{>} _{\sim}$10$^8$\solm, not correcting
for inclination effects.
Models resulting from a rotation curve including the effect of such a
central mass are shown in solid and dashed red contours.
Including a possible contribution from a compact nuclear stellar cluster
this value is consistent with a black hole mass of 1.7$\times$10$^7$\solm ~
estimated from nuclear H$_2$O-maser emissions by Greenhill \&
Gwinn (1997). 
}
\end{figure}

\begin{figure}
\psfig{file=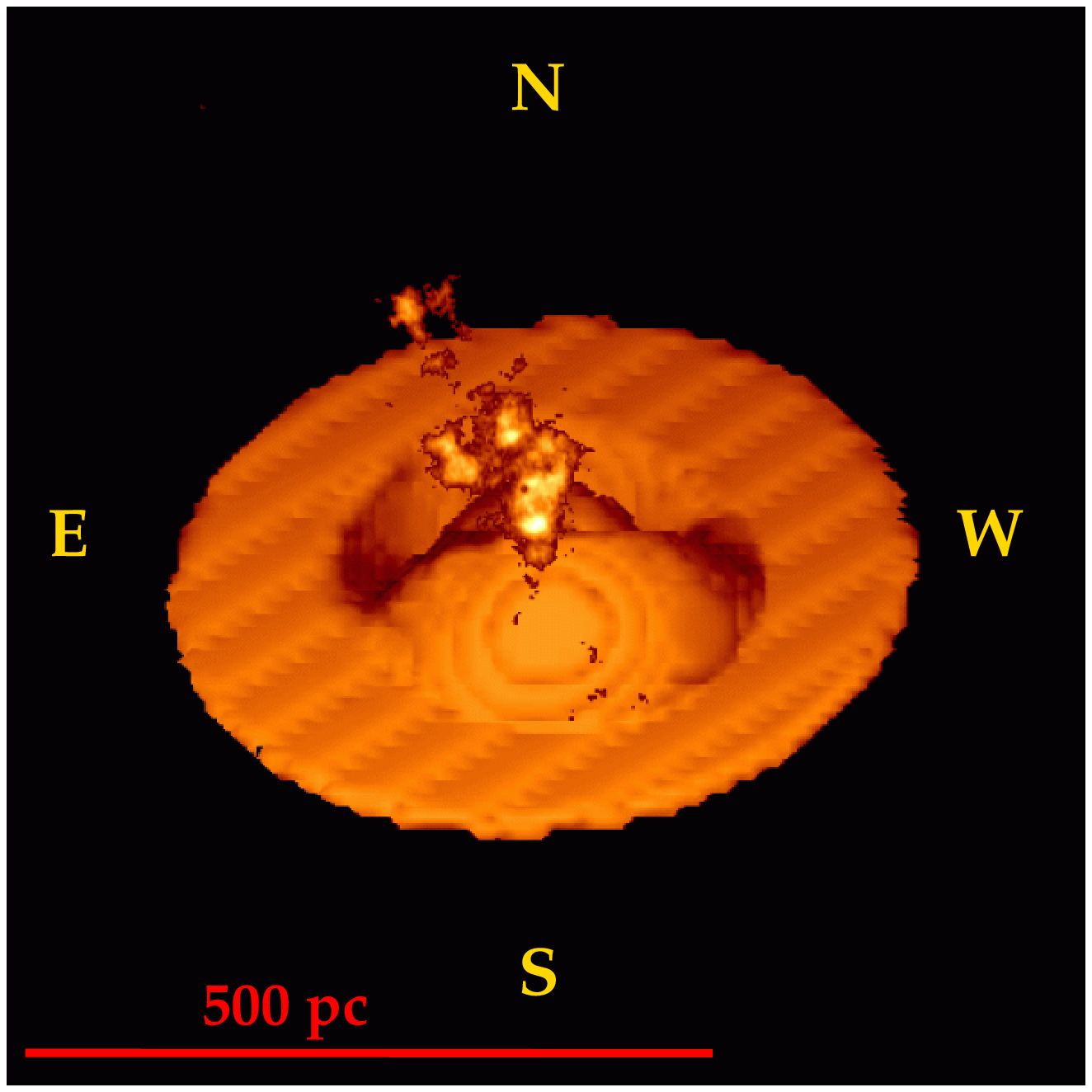,height=15cm,width=15.0cm,angle=0.0}

\figcaption[]{Spatial geometry of the warp model of NGC~1068. Bright
colors of the warped disk indicate sections that are closer to the
observer, darker colors those further away. 
The ionization cone as seen in the [O~III] line emission of Macchetto 
et al. (1994) is displayed as the brightest structure.
The three-dimensional geometry of the warp naturally provides a 
cavity for the ionization cone in consistency with the observed 
orientation.
}
\end{figure}


\clearpage
\begin{references}
\rf{Antonucci, R.R.J., Miller, J.S., 1985, \apjj {297}{621}}
\rf{Athanassoula, E., 1992, \mn{259}{345}}
\rf{Baker, A.J, Scoville, N.Z., 1998, Proceedings of IAU Symposium
    184, ed. Y. Sofue (Dordrecht:Kluwer), 221}
\rf{Binney, J., 1992, \arasa {30}{51}}
\rf{Bland-Hawthorn, J., Gallimore, J.F., Tacconi, L., Brinks, E.,
    Baum, S.A., Antonucci, R.R.J., Cecil, G.N., 1997, \appss {248}{9}}
\rf{Cameron, M., Storey, J.W.V., Rotaciuc, V., Genzel, R., Verstraete,
   L., Drapatz, S., Siebenmorgen, R., Lee, T.J., 1993, \apj {419}{136}}
\rf{Catchpole, R.M., Boksenberg, A., 1997, \appss {248}{79}}
\rf{Friedli, D., Martinet, L., 1993, \asa{277}{27}}
\rf{Garcia, A.M., 1993, \asas {100}{47}}
\rf{Greenhill, L.J., Gwinn, C.R., 1997, \appss{248}{261}}
\rf{Helfer, T.T., Blitz, L., 1995, \apjj {450}{90}}
\rf{Jackson, J.M., Paglione, T.A.D., Ishizuki, S., Nguyen-Q-Rieu,
    1993, \apjjl {418}{L13}}
\rf{Lumsden, S.L., Moore, T.J.T., Smith, C., Fujiyoshi, T.,
    Bland-Hawthorn, J., Ward, M.J., 1999, \mn {303}{209}}
\rf{Macchetto, F., Capetti, A., Sparks, W.B., Axon, D.J., Boksenberg,
   A., 1994, \apjjl {435}{L18}}
\rf{Malkan, M.A., Gorjian, V., Tam, R., 1998, \apjjs {117}{25}}
\rf{Meixner, M., Puchalsky, R., Blitz, L., Wright, M., Heckman, T.,
    1990, \apjj {354}{158}}
\rf{Miyoshi, M., Moran, J., Herrnstein, J., Greenhill, L., Nakai, N.,
   Diamond, P., Inoue, M., 1995, Nature, 373, 127}
\rf{Peterson, B., 1997, An Introduction to Active Galactic Nuclei,
    Cambridge, Cambridge Univ. Press}
\rf{Quillen, A.C., De Zeeuw, P.T., Phinney, E.S., Phillips,
   T.G., 1992, \apjj {391}{121}}
\rf{Osterbrock, D.E., 1977, \apjj {215}{733}}
\rf{Regan, M.W., Mulchaey, J.S., 1999, \ajj {in press}{}}
\rf{Salamanca, I. et al., 1994, \asa {282}{742}}
\rf{Schinnerer, E., Eckart, A., Tacconi, L.J., Genzel, R., 
    Downes, D. 1999, in preparation}
\rf{Schinnerer, E., Eckart, A., Tacconi, L.J.,  1999, in preparation}
\rf{Schinnerer, E., 1999 Ph.D. thesis, Ludwig-Maximilians-Universit\"at, 
    M\"unchen, Germany}
\rf{Schmitt, H.R., Kinney, A.L., 1996, \apj{463}{498}}
\rf{Scoville, N.Z., Matthews, K., Carico, D.P., Sanders, D.B., 1988, 
    \apjjl{327}{L61}}
\rf{Sellwood, J.A., Wilkinson, A., 1993, Rep. Prog. Phys., 56, 173}
\rf{Sparke, L.S., 1996, \apjj {473}{810}}
\rf{Tacconi, L.J., Gallimore, J.F., Genzel, R., Schinnerer, E.,
    Downes, D., 1997, \appss {248}{59}}
\rf{Tacconi, L.J., Genzel, R., Blietz, M., Cameron, M., Harris, A.I.,
    Madden, S., 1994, \apjjl{426}{L77}}
\rf{Telesco, C.M., Decher, R., 1988, \apjj {334}{573}}
\rf{Thatte, N., Quirrenbach, A., Genzel, R., Maiolino, R., Tecza, M., 1997,
    \apjj {490}{238}}
\rf{Tubbs, A.D., 1980, \apjj{241}{969}}
\rf{Young, S., Packham, C., Hough, J.H., und Efsathiou, A., 1996,
    \mn {283}{L1}}
\end{references}
\end{document}